\documentclass[prd, twocolumn, nofootinbib]{revtex4}

\usepackage{epsfig} 
\usepackage{amsmath}

\newcommand{\beq}{\begin{equation}}
\newcommand{\eeq}{\end{equation}}
\newcommand{\beqa}{\begin{eqnarray}}
\newcommand{\eeqa}{\end{eqnarray}}
\newcommand{\om}{\Omega_m}

\newcommand{\rfrw}{r_{\rm FRW}}
\newcommand{\kmin}{\kappa_{\rm min}}

\newcommand{\lang}{\langle} 
\newcommand{\rang}{\rangle} 
\newcommand{\sg}{\sigma_2} 
\newcommand{\sa}{\sigma_1} 
 
\newcommand{\ls}{\mathrel{\raise0.27ex\hbox{$<$}\kern-0.70em \lower0.71ex\hbox{{
$\scriptstyle \sim$}}}}

\begin{document} 

\title{Gravitational Wave Sirens as a Triple Probe of Dark Energy} 
\author{Eric V. Linder}
\affiliation{Berkeley Lab \& University of California, Berkeley, CA 94720, USA}

\date{\today}

\begin{abstract} 
Gravitational wave standard sirens have been considered as precision 
distance indicators to high redshift; however, at high redshift standard 
sirens or standard candles such as supernovae suffer from lensing noise.  
We investigate lensing noise as a signal instead and show how measurements 
of the maximum demagnification (minimum convergence) probe cosmology in a 
highly complementary manner to the distance itself.  Revisiting the 
original form for minimum convergence we quantify the bias arising from the 
commonly used approximation.  Furthermore, after presenting 
a new lensing probability function we discuss how the width of the lensed 
standard siren amplitude distribution also probes growth of structure.  Thus 
standard sirens and candles can serve as triple probes of dark energy, 
measuring both the cosmic expansion history and growth history.  
\end{abstract} 


\maketitle

\section{Introduction \label{sec:intro}}

Observations of distant standardized sources carry information not only 
on the global cosmology but the signal propagation through the intervening 
universe.  In particular, gravity affects the propagation on all scales, 
from the global curvature of spacetime to the local inhomogeneities of 
the mass distribution.  In this sense, detecting the acceleration of the 
cosmic expansion through measuring distances from the luminosity, size, 
or amplitude of a source is as much gravitational lensing as is the image 
distortion pattern around massive structures. 

While the global spacetime properties determine the distances in a manner 
fixed by the cosmology, inhomogeneities spread the derived distances 
about the ``truth'' or smooth universe value according to a lensing 
probability distribution function depending on the energy-momentum 
distribution and the convergence and shear it induces in the image.  
An interesting feature of this probability distribution is the presence 
of a minimum convergence or maximum demagnification corresponding 
to an ``empty beam'' \cite{dr1}.  This cutoff carries cosmological 
information. 

Thus the mean distance measured to standardized sources such as Type Ia 
supernovae in luminosity or inspiraling supermassive black hole binaries 
in gravitational wave amplitude (standard sirens \cite{holzhughes,dalal}) 
can probe cosmology, but so can the lensed amplitude distribution 
through the minimum convergence and the width of the distribution.  
Gravitational wave sources in particular may be measured to high 
redshift $z\gg1$ where lensing effects are quite evident.  Standard 
sirens and candles thus have the potential to be triple probes of cosmology. 

In \S\ref{sec:beam} we derive a general expression for the minimum 
convergence and maximum demagnification, correcting an often used 
expression in the literature.  Implications for 
gravitational wave standard sirens and supernova standard candles 
are examined in \S\ref{sec:stds}. 
The use of the minimum convergence as a cosmological probe in its own 
right is considered in \S\ref{sec:lever}, along with its complementarity 
with distance measurements, and the potential for the lensed distribution 
width to serve as a growth probe.  In the Appendix we briefly discuss 
effective averaging procedures in an inhomogeneous universe and why the 
standard weak gravitational lensing formula remains valid.

\section{Maximum Demagnification \label{sec:beam}} 

For signals on null geodesics such as light rays from supernovae or 
gravitational waves from inspiraling supermassive black hole binaries, 
the propagation through 
the universe can be described by the geometric optics approximation when 
the wavelength is much smaller than the scales of inhomogeneity.  This 
then leads to the optical scalar equations for a ray bundle \cite{sachs}, 
whose area defines a luminosity or amplitude distance.  

In a globally 
Friedmann universe, taking Ricci focusing, or convergence, to dominate 
over rotation and shear leads to the beam equation \cite{dr2,lin88} 
\beqa 
\ddot r&\!\!+\!\!&[3+q(z)](1+z)^{-1}\,\dot r \nonumber \\ 
&\!\!+\!\!&(3/2)(1+z)^{-2}\,r \,\sum_w (1+w)\alpha_w(z) \Omega_w(z)=0. 
\eeqa 
Here $r$ is the angular diameter distance (related to luminosity or 
amplitude distances through redshift factors $1+z$ in a Liouville or 
phase space density conserving system), an overdot denotes differentiation 
with respect to $z$, and $q(z)$ is the deceleration parameter.  Each 
energy-momentum component has a dimensionless energy density 
$\Omega_w(z)=8\pi\rho_w(a)/[3H^2(z)]$, where $H(z)$ is the Hubble 
parameter, a smoothness parameter $\alpha_w=(\rho_w)_{\rm smooth}/\rho_w$, 
and an equation of state, or pressure to energy density, parameter 
$w=p_w/\rho_w$.  

Note that only the global properties of the universe enter in the 
first two terms, while the local, inhomogeneous properties treated by 
the smoothness parameter $\alpha$ appear in the third term.  We can 
compare two universes with the same global dynamics but different local 
properties, or two signal paths within the same universe, to find \cite{lin88}
\beqa 
r(z_0,z)=\bar r(z_0,z)&{}&\!\!\!\!\!\!\!\!+\int_{1+z_0}^{1+z} 
du\,[uH(u)/H_0] 
[\bar Q(u)-Q(u)] \nonumber \\ 
&\ &\qquad \times\ r(1+z_0,u)\,\bar r(u,1+z), \label{eq:rbarr}
\eeqa 
where $Q(u)=(3/2)u^{-2}\sum_w (1+w)\alpha_w(u)\Omega_w(u)$.  Note that 
this is valid for the distance of some source at redshift $z$ from an 
object (either lens or observer) at redshift $z_0$.  
We will be particularly interested in comparing two signal paths in the 
same universe so the only difference between the barred distance and 
the unbarred one comes from different $\alpha$ along the lines of sight. 

If we now specialize to the case where there is no contribution 
of some component $x$ along the entire line of sight, i.e.\ $\alpha_x=0$, 
this represents the least gravitational focusing, or minimum convergence, 
possible with respect to the component $x$.  That is, one cannot take away 
more than there 
is\footnote{We do not consider $w<-1$, where Ricci focusing 
can actually go negative.  Also, shear acts to {\it increase\/} the Jacobian, 
i.e.\ amplification, so the convergence here is indeed the minimum.}.  
As shown 
in \cite{lin88} (also cf.\ \cite{schnw} for the matter only case), the 
distance $r$ is a monotonic function of $Q$.  Therefore the minimum 
$\alpha$ gives the greatest distance and faintest source, i.e.\ greatest 
deamplification relative to a smooth, Friedmann universe. 

The convergence is directly related to this distance ratio, since the 
distance came from considering the beam area.  Comparing a given line 
of sight to the dynamically equivalent smooth universe case (see the Appendix 
for further discussion), using Eq.~(\ref{eq:rbarr}) we have 
\beqa 
\kappa&\equiv&1-\frac{r_\alpha}{\rfrw} \label{eq:ktrue} \\ 
&=&-\frac{3}{2}\int_1^{1+z} \frac{dy}{y}\om(y)\frac{H(y)}{H_0} (1-\alpha) 
\nonumber \\ 
&\ &\qquad\qquad\qquad \times\ r_\alpha(y) 
\frac{\rfrw(y,1+z)}{\rfrw(1+z)} \label{eq:kralfmid} \\ 
&=&-\frac{3}{2}\om\,(1-\alpha)\int_1^{1+z} dy \frac{y^2}{H(y)/H_0}  
\nonumber \\ 
&\ &\qquad\qquad\qquad \times\ r_\alpha(y) \frac{\rfrw(y,1+z)}{\rfrw(1+z)} 
\,, \label{eq:kralfbot} 
\eeqa
where in the second line we assume for simplicity that only one component 
is not smooth (e.g.\ dark energy is smooth but matter can clump) and in 
the last line we assume 
$\alpha$ is constant with redshift.  These simplifications are not 
necessary, but for the minimum convergence $\alpha=0$, which is indeed 
constant (see the Appendix for the general case). 

The triplet of distances in Eqs.~(\ref{eq:kralfmid}), (\ref{eq:kralfbot}) 
is often written as $r_l r_{ls}/r_s$, 
representing the lens distance, lens-source distance, and source 
distance.  Note that a common practice appears to be to use FRW 
distances for all three quantities, even when calculating the minimum 
convergence $\kmin$.  As we see following \cite{lin88}, the proper 
expression derived from the optical scalar equations actually 
distinguishes between the types of distances.  

We now give another brief, nonrigorous motivation from the ray deflection 
equation.  (See \cite{schnef} for more details.) 
The position of an image that is gravitationally deflected by an angle 
$\beta$ is 
\beq 
\vec\theta \to \vec\theta-\frac{r_{ls}}{r_s}\vec\beta \,.\label{eq:lenseq} 
\eeq 
Here the distances are purely 
geometric and the ray position is integrated along the unperturbed 
path, so these distances should correspond to the global spacetime, 
i.e.\ be FRW distances.  The Jacobian of the transformation~(\ref{eq:lenseq}) 
gives the convergence and involves 
\beq 
\frac{\partial\beta_i}{\partial\theta_j}=\frac{\partial\beta_i}{\partial 
x_j}r_l \to r_l \frac{\partial^2\phi}{\partial x_i\partial x_j}, 
\eeq 
where $\phi$ is the gravitational potential of the mass perturbation. 
Because these quantities specifically have to do with the local, 
inhomogeneous potential, the distance here is not the FRW distance but 
$r_\alpha$.  While this may possibly be more an intuitive rationale, 
we see that both approaches give the same result. 

Although using all FRW distances is formally incorrect, under most 
circumstances the common expression 
\beq 
\kappa_{\rm approx}=-\frac{3}{2}\om\int_1^{1+z} dy \frac{y^2}{H(y)/H_0}  
\frac{r_l^{\rm FRW} r_{ls}^{\rm FRW}}{r_s^{\rm FRW}} \label{eq:kusual}
\eeq 
is an excellent approximation, as discussed in the Appendix.  The 
notable exception is when calculating $\kmin$, that is Eq.~(\ref{eq:ktrue}), 
(\ref{eq:kralfmid}), or (\ref{eq:kralfbot}) with $\alpha=0$.  
Here one explicitly 
deviates from FRW by requiring that the ray bundle passes through emptied 
space.  Of course this relies on the validity of the Dyer-Roeder approach 
to light propagation, where the global dynamics is FRW and only Ricci 
focusing is important.  We will be most interested here in the minimum 
convergence, and as stated, by neglecting shear we obtain a true lower 
limit.  Equation~(\ref{eq:ktrue}), which follows from the optical beam 
equation and flux conservation in this case \cite{wbg76}, then directly 
leads to Eq.~(\ref{eq:kralfbot}).  
The next section examines the implications of the breakdown of the 
common formula.

\section{Lensing of Standard Candles and Sirens \label{sec:stds}} 

\begin{figure}[!htb]
\begin{center} 
\psfig{file=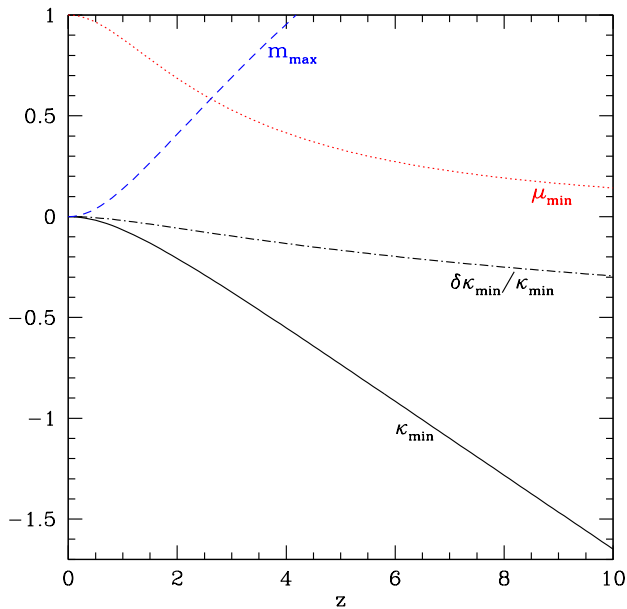,width=3.4in} 
\caption{Lensing causes appreciable effects in standard candle/siren 
amplitudes, especially at high redshifts.  Here the true minimum 
convergence $\kappa_{\rm min}$, minimum amplitude $\mu_{\rm min}$ (as 
a fraction of the true, standard amplitude), and maximum demagnification 
$m_{\rm max}$ are plotted vs.~redshift, along with the fractional error 
$\delta\kmin/\kmin$ from using the approximate rather than true expression. 
}
\label{fig:kminvsz} 
\end{center} 
\end{figure}

Convergence can also be expressed in terms 
of the amplification $\mu$, where the standard amplitude of the candle or 
siren is taken to be unity, or the demagnification $m$ by 
\beqa 
\mu&=&(1-\kappa)^{-2} \\ 
m&=&5\,\log (1-\kappa) \,.
\eeqa 
These expressions do not assume weak lensing but do neglect Weyl shear. 
The minimum convergence $\kmin$ occurs when all the energy density that 
can be emptied from the beam, i.e.\ can clump elsewhere, has been so 
$\alpha=0$ in Eq.~(\ref{eq:ktrue}). 
Figure~\ref{fig:kminvsz} shows the values of $\kmin$, and the associated 
$\mu_{\rm min}$ and $m_{\rm max}$, as a function of redshift, within a 
flat LCDM cosmology with $\om=0.3$.  

By $z=3$ 
one has $\kmin=-0.37$ (by definition $\kmin$ is negative) and a 
minimum amplification of 53\% (the standard candle/siren can appear 
53\% of its true amplitude) or maximum demagnification of 0.69 mag.  For 
a standardized candle with, say, 0.15 mag dispersion, this is a significant 
alteration so lensing can have a substantial cosmological impact.  A fitting 
form for $\kmin$, accurate to better than 0.01 mag out to $z=10$, is 
\beq 
\kmin\approx 
\begin{cases} 
-0.067\,z^{1.58} & z<3, \\ 
-0.374-0.182(z-3) & z\ge 3. 
\end{cases} 
\eeq

\begin{figure}[!htb]
\begin{center} 
\psfig{file=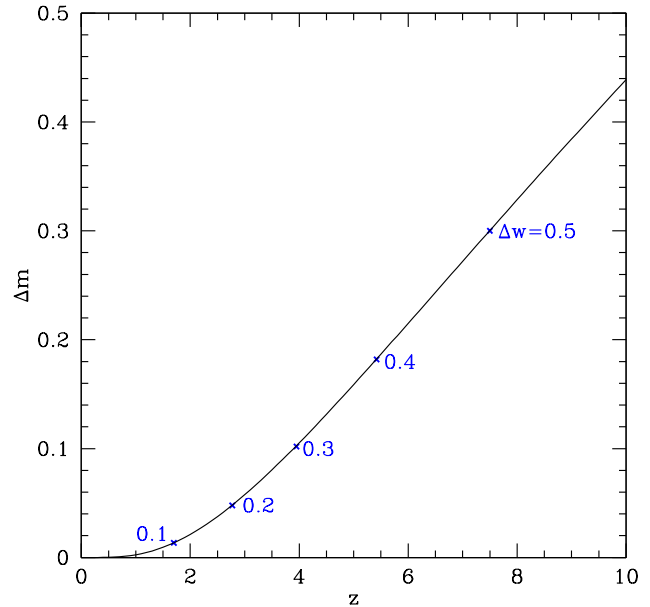,width=3.4in} 
\caption{Magnitude error from using the approximate rather 
than the true minimum convergence.  Such an error leads to a bias in 
the derived cosmology; labels along the curve show the equivalent 
offset $\Delta w$ in a constant equation of state, relative to the 
true value of $w=-1$, caused by using the approximate expression at 
redshift $z$. 
}
\label{fig:kminasw} 
\end{center} 
\end{figure}

Also of importance is the error caused by using the common approximation 
of Eq.~(\ref{eq:kusual}).  The difference between the correct and 
approximate expressions for the convergence grows with redshift as shown 
by the dot-dashed line in Fig.~\ref{fig:kminvsz}.  
This reaches 10\% misestimation of $\kmin$ 
at $z=3$, which is equivalent to a shift in distance scale from 
misestimating the dark energy equation of state by $\Delta w=0.22$.  That 
constitutes a substantial bias for precision probes of cosmology so 
the exact expression for convergence, Eq.~(\ref{eq:ktrue}), must be used 
for accurate cosmology utilizing the minimum convergence.  
Figure~\ref{fig:kminasw} plots the error in maximum demagnification 
arising from using the approximate expression vs.\ redshift and shows 
the equivalent bias that would be induced in $w$.  One might also worry 
that since 
$\kmin$ enters into the universality scaling of the lensing probability 
distribution function, an error in $\kmin$ propagates into all lensing 
distribution quantities that are derived from the universal distribution 
via that scaling.  However a consistent calculation of lensing 
distribution functions by, e.g., ray tracing simulations will avoid this.

In the next section we investigate the use of the minimum convergence 
as a source of cosmological information, but first we emphasize that 
statistically lensing does not affect the mean amplitude measured for 
the standard candles or sirens.  Thus, given sufficient numbers of 
measurements their role as distance indicators remains intact (see 
\cite{holzlin05} for detailed discussion).  As we 
now see, however, substantial information also exists in the distribution 
of lensed amplitudes, if well sampled.

\section{Cosmological Leverage \label{sec:lever}} 

\subsection{Observational Issues \label{sec:obs}} 

Measurement of the minimum convergence requires sufficient sources 
to map out the low amplification end of the lensing probability 
function.  Furthermore, what is actually observed is the convolution 
of the imperfectly standardized luminosity or strain amplitude function 
of the sources with the lensing amplification.  In addition, astrophysical 
effects such as dust extinction can also dim the luminosity, making it 
difficult to measure precisely the minimum convergence from the lower 
limit of the lensing amplification.  We briefly discuss these issues 
(also see \cite{snapping}) but a detailed treatment awaits future work. 

To measure accurately the cutoff $\kmin$ one must observe many standard 
sources to find the lower limit to their amplitudes, $\mu_{\rm min}$. 
At high redshift the number of measured sources available will involve the 
source volume density and rate and the detection threshold for increasingly 
distant objects.  Since gravitational wave amplitudes diminish linearly 
with distance, rather than going as the inverse square like supernova 
luminosities, one might expect to detect standard sirens to greater 
distances.  However, observations of their electromagnetic counterparts 
necessary to measure redshifts might become more difficult.  Of course 
in seeking to measure the most deamplified sources at a given redshift we 
make the detection more challenging and must ensure Malmquist bias does 
not arise.  A well designed survey can ameliorate these concerns. 

If the sources have tightly standardized amplitudes, either luminosity in 
the case of a standardized candle or metric strain in the case of a 
standardized siren, then for $1-\mu_{\rm min}$ much larger than the 
residual intrinsic dispersion, the value of $\kmin$ should be measurable.  
We see from Fig.~\ref{fig:kminvsz} that by $z=2$ the minimum 
convergence corresponds to a demagnification of 0.41 mag, much larger than 
well standardized source dispersions.  Thus, $\kmin$ for higher redshifts 
should lie several sigma out on the intrinsic source amplitude probability 
tail, reducing confusion of source properties with lensing effects.  
(However, we may not know the source amplitude function well so far from 
its mean so it is possible that using Gaussian statistics for it will give 
an overoptimistic assessment.)  Since the best gravitational wave sirens 
can be standardized on an individual basis more precisely than individual 
supernovae, this should be less of a problem for sirens, though their 
statistics for mapping the lensing distribution may be poorer.  They are also 
unaffected by residuals to dust extinction corrections that could confuse 
the interpretation of lensing deamplification. 

An important point concerns whether truly empty beams exist in a practical 
sense.  Investigation requires ray tracing through large N-body simulations, 
convolved with details of survey strategy, and is beyond the scope of this 
article.  However, to develop the idealized promise outlined by the 
following results, this will need to be done.  
Two aspects seem hopeful: One is that the sharp cutoff of the 
magnification probability on the low end, discussed in \S\ref{sec:lensdisp}, 
means that there may be relatively little error in characterizing $\kmin$ 
even if one measures where the distribution goes to, say, 10\% of maximum 
amplitude rather than 0\%.  
The other is the possibility that while truly empty beams are rare, 
there may be a relation between the effective observed $\kmin^{\rm obs}$ 
and the true $\kmin$; if this can be calibrated through N-body 
simulations then one can still use the technique, albeit at reduced 
accuracy\footnote{I am particularly grateful to Daniel Holz for this idea}.

\subsection{Minimum Amplification as Expansion Probe \label{sec:exp}}

\begin{figure}[!htb]
\begin{center} 
\psfig{file=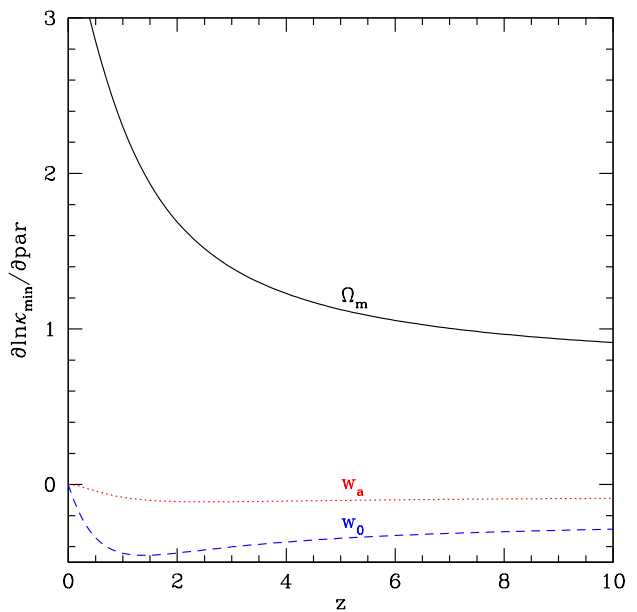,width=3.4in} 
\caption{The Fisher derivative of the logarithm of the observable $\kmin$ 
with respect to the cosmological parameters indicates the sensitivity 
to the parameters at different redshifts.  Similar shapes of curves over 
a redshift range indicate degeneracy between parameters and relatively 
little additional information, here occurring for $z>5$. 
}
\label{fig:sens} 
\end{center} 
\end{figure}

Having noted these issues, we approximate them with a highly simplified 
error model and examine the cosmological information in the minimum 
convergence, leaving detailed experiment design to the future.  Employing 
the Fisher matrix formalism we assess how measurements of $\kmin$ at 
various redshifts, through either standard sirens or candles, propagate 
into cosmological parameter constraints.  Figure~\ref{fig:sens} illustrates 
the sensitivity to the matter density $\om$ and the dark energy parameters 
$w_0$ and $w_a$, with the standard time dependent equation of state 
$w(a)=w_0+w_a(1-a)$, for a spatially flat, $\om=0.28$ LCDM universe.

The main sensitivity is to the matter density, at $z=1.5$ some 4 times more 
than to $w_0$ and 19 times more than to $w_a$.   At high redshift the 
sensitivity declines and the curves become similar in shape, indicating 
substantial 
degeneracy.  Thus we do not anticipate that sources at $z\gtrsim5$ will 
contribute effectively to cosmological model estimation.  Since such 
distant sources would be difficult to observe accurately and in 
sufficiently large numbers (and lines of sight possessing the minimum 
convergence, i.e.\ no matter along the entire path, become increasingly 
rare), it is fortunate that the sensitivity indicates no need for such 
deep surveys. 

Proceeding further, we investigate the cosmological leverage of $\kmin$ 
as a probe in isolation.  Considering measurements of 10\% precision, 
even 100 data points, every 0.1 in redshift over $z=0.1-10$, do not 
put tight constraints on dark energy properties; the uncertainty on 
$w_0$ is 0.36 and on $w_a$ is 1.  However, this is due to a strong 
degeneracy among the parameters and if we consider combining $\kmin$ 
measurements with the distance measurements of the standard sirens or 
candles themselves, the picture is quite different. 

\begin{figure}[!htb]
\begin{center} 
\psfig{file=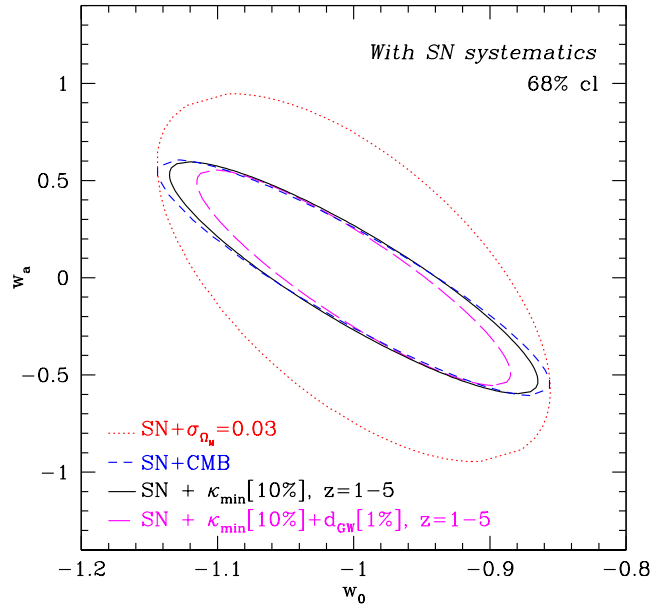,width=3.4in} 
\caption{Minimum amplification has good complementarity with distance 
probes for determining the dark energy equation of state, shown here as 
the $1\sigma$ joint confidence contours in $w_0$, $w_a$.  It is closely 
equivalent in its constraint leverage to CMB data, but without extrapolating 
to the early universe. 
}
\label{fig:kminell} 
\end{center} 
\end{figure}

The minimum convergences have strong complementarity with the distances 
and improve the distance constraints substantially.  
Figure~\ref{fig:kminell} shows that as a complement to supernova (SN) 
distances out to $z=1.7$ of the quality expected from the SNAP supernova 
survey, the minimum convergence has essentially the impact of the Planck 
cosmic microwave background measurements.  Here we considered 10\% 
measurements of $\kmin$ every 0.5 in redshift over $z=1-5$. 
Note that this may be a reasonable redshift range as for lower redshifts 
lensing effects are weaker and harder to separate from the standard 
amplitude dispersion, while for higher redshifts the sources may be 
harder to detect in sufficient numbers to measure robustly the minimum 
amplification cutoff.

In terms of the parameter uncertainties, adding the $\kmin$ measurements 
tightens constraints on $w_a$ by a factor 1.5 and decreases the contour 
area by a factor 2.7.  An advantage of using $\kmin$ is that there is 
no need to extrapolate the behavior of the dark energy to $z>1000$, as 
there is for using the CMB constraint.  If one does add CMB data as well, 
then the $w_a$ uncertainty decreases by a further 7\% and the area shrinks 
by a further 29\%. 

Using the distances measured from the gravitational wave (GW) sirens 
themselves, rather than from supernovae, does not give as strong 
constraints if we assume that their 
accuracy is limited to 1\% over the same redshift range as we use $\kmin$ 
-- remember that lensing degrades the precision of these high redshift 
distance determinations.  Use of GW distances and $\kmin$ gives contours 
outside the range shown in 
Fig.~\ref{fig:kminell} (though lower redshift GW distances would tighten 
the constraints).  Adding GW distances to the previous set of SN distances 
and $\kmin$ improves $w_0$ estimation by 15\%, $w_a$ by 7\% and the area 
shrinks by 20\%, as shown in Fig.~\ref{fig:kminell}.  Thus, SN distances 
(or low redshift GW distances) are important complements.

\subsection{Lensing Dispersion as Growth Probe \label{sec:lensdisp}} 

Note that both the standard siren/candle distances and the minimum 
amplification of the lensing probability function measured through 
the siren/candle amplitude distribution are geometric probes of the 
cosmic expansion history.  However, the lensing itself arises from 
mass inhomogeneities and so contains information on the growth history 
of structure in the universe.  This last use has been discussed by, 
e.g., \cite{frie96,seljakholz,dodelval} as a means of measuring the current 
strength of mass fluctuations, $\sigma_8$.  Basically, the width of 
the lensing induced dispersion is a weighted function of the mass 
power spectrum. 

Here we briefly outline how measurements of the lensing dispersion 
as a function of redshift probe the linear density growth factor, 
$g(z)=(1+z)\,[\delta\rho_m/\rho_m](z)$.  The mass power spectrum is 
proportional to the linear growth factor squared and the transfer 
function.  A standard approximation, which will need to be improved 
for future precision cosmology, is that the transfer function is 
insensitive to deviations from the LCDM model; thus the lensing dispersion 
may allow determination of the linear growth (with no galaxy bias 
entering).  If successful, this means 
that by measuring 
distances through standard sirens/candles, one actually has three probes 
of cosmology: 1) the mean amplitude gives a geometric measure of the 
distance, 2) the minimum amplification gives a complementary geometric 
measure consisting of distance ratios, and 3) the dispersion encodes 
the growth history through the linear growth factor.  

Defining the width or dispersion of the lensing probability distribution 
function (pdf) is nontrivial.  In an effort to develop understanding of 
lensing pdf's suitable for the high redshifts and precisions needed for 
next generation surveys, \cite{holzlin08} has investigated the 
hyperLandau (HL) form 
\beq 
P(\mu)=\frac{2^{-s}}{\sg\Gamma(s)}\, e^{-\frac{1}{2}[\frac{\mu-\mu_\star}{\sa}+
e^{-\frac{\mu-\mu_\star}{\sg}}]} \label{eq:hlfit}
\eeq 
with some success.  This is a three (or fewer) parameter form, with $\sa$, 
$\sg$, $\mu_\star$, where $\Gamma(s)$ is the gamma function and 
$s=\sg/(2\sa)$.  The variance is proportional to $\sg^2$ and when $s$ 
is independent of redshift 
then the width of the distribution as a function of redshift scales as 
$\sg(z)$.  This can be calibrated against ray tracing simulations and may 
allow the growth factor $g(z)$, to which it should be proportional, to 
be extracted, though further work is required to verify 
this\footnote{Note 
that one may actually use the linearly scaled variable 
$u=(\mu-\mu_{\rm min})/(1-\mu_{\rm min})$ to improve the universality, 
i.e.\ cosmology independence, of the pdf \cite{munshijain}.  Also, the 
steep dropoff of the form (\ref{eq:hlfit}) on the deamplification side 
gives support to $\kmin$ being reasonably precisely measurable.}. 

\begin{figure}[!htb]
\begin{center} 
\psfig{file=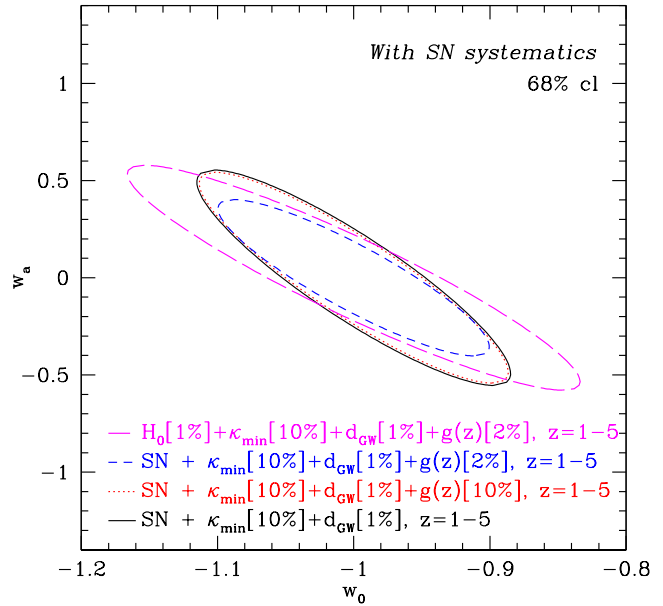,width=3.4in} 
\caption{Standard siren or candle measurements also probe the growth 
history through the width of the lensed amplitude distribution.  Accuracy 
approaching 2\% on the linear growth factor is required for significant 
additional constraints on the dark energy equation of state, but less 
stringent growth estimation may provide important crosschecks or tests 
of extended gravity.  Here the solid curve corresponds to the long-dashed 
curve of Fig.~\ref{fig:kminell}, without growth information. 
}
\label{fig:kminellgro} 
\end{center} 
\end{figure}

If this does allow measurement of $g(z)$, then GW standard sirens or 
SN standard candles provide a triple probe of dark energy, including 
both the expansion and the growth histories.  Figure~\ref{fig:kminellgro} 
shows that 10\% measurements of the growth factor $g(z)$ over the same 
redshift range as $\kmin$ does not add much knowledge in $w_0$, $w_a$ -- 
but recall that one of the points of combining expansion probes with 
growth probes is to provide a crosscheck and a window on gravity beyond 
Einstein, e.g.\ through the gravitational growth index $\gamma$ 
\cite{hutlin06}, not just $w_0$, $w_a$.  

In order to see significant improvements in the 
$w_0$-$w_a$ plane, one requires 2\% accuracy in the $g(z)$ measurements 
from the width of the lensed amplitude distribution, which may be 
challenging.  If achievable, this delivers 12\% improvement in 
determining $w_0$, 27\% in $w_a$, and a factor 1.51 in the contour area. 
From Fig.~\ref{fig:kminellgro} we see that we also require 2\% accuracy 
in measuring $g(z)$ to roughly substitute for supernovae distances, 
i.e.\ to have a purely gravitational wave standard siren constraint, 
though this also requires either precision low redshift distance or 
Hubble constant measurements.

\section{Conclusions \label{sec:concl}} 

Standardized siren or candle measurements of the mean distance-redshift 
relation provide precise cosmological parameter estimation, but their 
full information is even richer.  The distribution in the binary 
black hole gravitational wave strain amplitude or supernova luminosity 
carries with it characteristics reflecting the propagation along the 
line of sight.  In particular, this distribution has a physically 
precise lower boundary due to the minimum convergence, which involves 
a distance ratio, providing a further geometric measurement of the 
cosmology. 

From the full, long known definition of the minimum convergence or maximum 
demagnification we show that a common approximation induces a substantial 
bias as measurements move toward higher redshifts and better precision. 
We explore several observational issues with using minimum amplification 
as a practical cosmological probe, suggesting that gravitational wave 
standard sirens might enable this method if sufficient numbers of sirens 
can be detected with precise redshifts. 

Along with a new probability distribution function for lensing amplification, 
we discuss how the width of the distribution might be used as a measure 
of the mass power spectrum or linear growth factor.  From a Fisher analysis 
it appears that the method would need to be developed to allow 2\% 
accuracy on the lensing dispersion for a significant gain in cosmological 
constraints, but weaker precision could still provide useful crosschecks 
or consistency tests of gravity. 

Substantial work needs to be done to realize the promises of the lensing 
distribution information for precision cosmology.  In particular, 
large ray tracing simulations are needed to verify and accurately calibrate 
the minimum detectable convergence and distribution width. 

While challenging, nevertheless it is exciting that distance measurements 
of gravitational wave standard sirens and possibly supernova standardized 
candles can act as a triple probe of cosmology and dark energy:   
1) the mean amplitude (unaffected by lensing) gives a geometric measure of the 
distance, 2) the minimum amplification gives a complementary geometric 
measure consisting of distance ratios, and 3) the dispersion encodes 
the growth history through the linear growth factor.  
If this potential can be realized, 
distance-redshift measurements of sirens or candles can provide probes 
of both the cosmic expansion and growth histories.

\acknowledgments

I thank Daniel Holz for useful discussions and the T6 group at Los Alamos 
National Laboratory for hospitality.  This work has been supported in part 
by the Director, Office of Science, Department of Energy under grant 
DE-AC02-05CH11231. 

\appendix 

\section{Averaging Inhomogeneity \label{sec:avg}} 

One might wonder whether the commonly used approximation 
Eq.~(\ref{eq:kusual}) is sufficiently accurate for use along arbitrary 
lines of sight, in comparison to the expression for arbitrary convergence 
$\kappa$, Eq.~(\ref{eq:ktrue}). 
This is an important point since $\kappa$ plays a central 
role in all gravitational lensing calculations.  The answer, fortunately, 
is that except for the special situation when one considers the particular 
lines of sight emptied of matter -- i.e.\ the minimum convergence treated 
in the main text -- use of the FRW distances is an excellent approximation 
and standard lensing formulas 
hold\footnote{In the {\it weak\/} lensing case specifically this holds 
nearly trivially as then $\kappa\ll1$ and so $\alpha\approx1$ 
(cf.~Eq.~\ref{eq:kralfmid}), so $r_\alpha\approx\rfrw$.}. 

Formally, one might write Eq.~(\ref{eq:kralfmid}) as 
\beq 
\kappa=\frac{3}{2}\om\int_1^{1+z} dy\,\frac{y^2}{H(y)/H_0}\,\delta(y) 
\, r_\alpha(y) \frac{\rfrw(y,1+z)}{\rfrw(1+z)}\,,
\eeq 
nearly the usual gravitational lensing form, except that instead of 
$\rfrw(y)$ we have $r_\alpha$ evaluated for $\alpha=\lang 
1+\delta(y'<y)\rang$, 
where $\delta=\delta\rho_m/\rho_m$. 
That is, $r(y)$ depends on the energy-momentum density within the beam at 
every redshift between source and observer, giving rise to a weighted 
smoothness parameter. 

This is basically a question of line of sight depth averaging over 
inhomogeneities 
to recover a FRW universe and is treated in detail in \cite{lin98} (along 
with other types of averaging and their possibly different results). 
Summarizing, light propagating along a random line of sight 
will experience both underdense and overdense regions.  For a typical 
inhomogeneity coherence scale $l$, there will be $L/l$ regions over a path 
length $L$ and the deviations from FRW behavior diminish with depth.  
This is made more precise in \cite{lin98} where it is shown that the $Q$ 
terms in Eq.~(\ref{eq:rbarr}) correspond to $\nabla^2\phi$, where $\phi$ 
is the gravitational potential.  Then for $\phi\sim{\mathcal O}(\epsilon^2)$, 
the correction to distances, $r_{\lang\alpha\rang}-\rfrw$, is 
$\sim\epsilon^2/(l/L)$, which must be much less 
than one for a globally FRW universe (see \cite{jlw}).  

The exception to this is when considering special lines of sight such 
that $\delta$ is constrained, e.g.~$\delta=-1$ for the minimum convergence. 
However, lensing generically involves the natural fluctuations of 
density along the line of sight and thus can robustly use FRW distances.


\begin{thebibliography}{99}

\bibitem{dr1}
C.C. Dyer \& R.C. Roeder, Ap. J. Lett 174, L115 (1972)

\bibitem{holzhughes} 
D.E. Holz \& S.A. Hughes, Ap. J. 629, 15 (2005) [arXiv:astro-ph/0504616] 

\bibitem{dalal} 
N. Dalal, D.E. Holz, S.A. Hughes, B. Jain, Phys. Rev. D 74, 063006 (2006) 
[arXiv:astro-ph/0601275] 

\bibitem{sachs} 
R.K. Sachs, Proc. Roy. Soc. London A 264, 309 (1961)

\bibitem{dr2}
C.C. Dyer \& R.C. Roeder, Ap. J. Lett 180, L31 (1973)

\bibitem{lin88}
E.V. Linder, A\&A 206, 190 (1988)

\bibitem{schnw} 
P. Schneider \& A. Weiss, Ap. J. 327, 526 (1988)

\bibitem{schnef} 
P. Schneider, J. Ehlers, E.E. Falco, Gravitational Lenses (Springer, 1992)

\bibitem{wbg76}
S. Weinberg, Ap. J. Lett. 208, L1 (1976)

\bibitem{holzlin05}
D.E. Holz \& E.V. Linder, Ap. J. 631, 678 (2005) [arXiv:astro-ph/0412173]

\bibitem{snapping} 
G. Aldering et al., Astropart. Phys. 27, 313 (2007) [arXiv:astro-ph/0607030] 

\bibitem{frie96}
J.A. Frieman, Commun. Astrop. 18, 323 (1996) [arXiv:astro-ph/9608068]

\bibitem{seljakholz} 
U. Seljak \& D.E. Holz, A\&A 351, L10 (1999) [arXiv:astro-ph/9910482]

\bibitem{dodelval}
S. Dodelson \& A.\ Vallinotto, Phys. Rev. D 74, 063515 (2006) 
[arXiv:astro-ph/0511086] 

\bibitem{holzlin08}
D.E. Holz \& E.V. Linder, in preparation

\bibitem{munshijain}
D. Munshi \& B. Jain, MNRAS 318, 109 (2000) [arXiv:astro-ph/9911502]

\bibitem{hutlin06}
D. Huterer \& E.V. Linder, Phys. Rev. D 75, 023519 (2007) 
[arXiv:astro-ph/0608681] 

\bibitem{lin98}
E.V. Linder, arXiv:astro-ph/9801122

\bibitem{jlw}
M.W. Jacobs, E.V. Linder, R.V. Wagoner, Phys. Rev. D 45, 3293 (1992); 
M.W. Jacobs, E.V. Linder, R.V. Wagoner, Phys. Rev. D 48, 4623 (1993) 


\end{thebibliography}
\end{document}